\documentclass[prc,aps,amssymb,twocolumn,showpacs]{revtex4}
\usepackage{amsfonts}
\usepackage{amsmath}
\usepackage{graphicx}
\usepackage{color}

\newcommand{\CPC}[3]{Comp.\ Phys.\ Comm.\ {\bf #1},\ #2 (#3)}

\newcommand{\IJMPE}[3]{Int.\ J.\ Mod.\ Phys.\ {\bf E#1},\ #2 (#3)}

\newcommand{\JETPL}[3]{JETP.\ Lett.\ {\bf #1},\ #2 (#3)}
\newcommand{\JPCS}[3]{J.\ Phys.\ Chem.\ Solids\ {\bf #1},\ #2 (#3)}

\newcommand{\NPA}[3]{Nucl.\ Phys.\ {\bf A#1},\ #2 (#3)}

\newcommand{\NJP}[3]{New\ J.\ Phys.\ {\bf #1},\ #2 (#3)}

\newcommand{\SPJETP}[3]{Sov.\ Phys.\ JETP\ {\bf #1},\ #2 (#3)}
\newcommand{\PAN}[3]{Phys.\ Atom.\ Nucl.\ {\bf #1},\ #2 (#3)}

\newcommand{\PRL}[3]{Phys.\ Rev.\ Lett.\ {\bf #1},\ #2 (#3)}

\newcommand{\PRA}[3]{Phys.\ Rev.\ {\bf A#1},\ #2 (#3)}

\newcommand{\PRC}[3]{Phys.\ Rev.\ {\bf C#1},\ #2 (#3)}

\newcommand{\PPNP}[3]
{Prog.\ Part.\ Nucl.\ Phys.\ {\textbf #1},\ #2 (#3)}

\newcommand{\ibid}[3]{{\bf #1},\ #2 (#3)}
\renewcommand\a{\alpha}
\renewcommand\b{\beta}
\renewcommand\d{\delta}

\renewcommand\r{\rho}

\renewcommand\t{\tau}

\renewcommand\c{\chi}
\renewcommand\j{\psi}
\renewcommand\o{\omega}
\newcommand\e{\epsilon}
\newcommand\g{\gamma}

\newcommand\m{\mu}
\newcommand\n{\nu}
\newcommand\x{\xi}
\newcommand\p{\pi}

\newcommand\s{\sigma}

\newcommand\w{\eta}

\renewcommand\P{\Pi}

\renewcommand\O{\Omega}

\newcommand\D{\Delta}

\newcommand{\fig}[1]{Fig.\,\ref{#1}}
\newcommand{\eq}[1]{Eq.(\ref{#1})}

\newcommand\lb{\left(}
\newcommand\rb{\right)}
\newcommand\ls{\left[}
\newcommand\rs{\right]}

\newcommand{\lan}{\langle}
\newcommand{\ran}{\rangle}
\newcommand\ua{\uparrow}
\newcommand\da{\downarrow}
\newcommand\ra{\rightarrow}

\newcommand{\non}{\nonumber\\}
\newcommand\pt{\partial}

\newcommand{\Tr}{{\rm Tr}}

\newcommand{\cl}{{\cal L}}

\newcommand{\bx}{{\mathbf x}}
\newcommand{\br}{{\mathbf r}}

\newcommand{\bk}{{\mathbf k}}
\newcommand{\bq}{{\mathbf q}}

\newcommand{\cf}{{\cal F}}
\newcommand{\idp}[2]{\int\frac{d^{\,#1}#2}{(2\p)^#1}}
\newcommand{\ie}{\emph{i.e.}}

\newcommand{\etal}{\emph{et al.}}
\begin{document}

\title{BCS-BEC Crossover and Thermodynamics in Asymmetric Nuclear Matter with Pairings in Isospin $I=0$ and $I=1$ Channels}
\author{\normalsize{Shijun Mao$^1$, Xuguang Huang$^{1,2}$ and Pengfei Zhuang$^1$}}
\affiliation{$^1$Physics Department, Tsinghua University, Beijing
100084, China\\ $^2$Frankfurt Institute for Advanced Studies and
Institute for Theoretical Physics, Frankfurt University, Frankfurt
am Main D-60438, Germany}

\date{\today}

\begin{abstract}
The BCS-BEC crossover and phase diagram for asymmetric nuclear
superfluid with pairings in isospin $I=0$ and $I=1$ channels are
investigated at mean field level, by using a density dependent
nucleon-nucleon potential. Induced by the in-medium nucleon mass
and density dependent coupling constants, neutron-proton Cooper
pairs could be in BEC state at sufficiently low density, but there
is no chance for the BEC formation of neutron-neutron and
proton-proton pairs at any density and asymmetry. We calculate the
phase diagram in asymmetry-temperature plane for weakly
interacting nuclear superfluid, and find that including the $I=1$
channel changes significantly the phase structure at low
temperature. There appears a new phase with both $I=0$ and $I=1$
pairings at low temperature and low asymmetry, and the gapless
state in any phase with $I=1$ pairing is washed out and all
excited nucleons are fully gapped.
\end{abstract}

\maketitle

%%%%%%%%%%%%%%%%%%%%%%%%%%%%%%%%%%%%%%%%%%%%%%%%%%%%%%%%%%%%%%%%%%%%%%%
\section {Introduction}\label{1}
%%%%%%%%%%%%%%%%%%%%%%%%%%%%%%%%%%%%%%%%%%%%%%%%%%%%%%%%%%%%%%%%%%%%%%%
As it is well-known, due to the condensate of nucleon-nucleon (NN)
Cooper pairs at sufficiently low temperature, a nucleon many-body
system, such as a large N nucleus or bulk nuclear matter in
neutron stars, will be in superfluid state with many various
interesting phenomena, like the properties of medium-mass
$N\approx Z$ nuclei produced at the radioactive nuclear beam
facilities~\cite{Goodman:1998,Goodman:1999}, the deuteron
formation in medium-energy heavy ion collisions~\cite{Baldo:1995},
and the equation of state of neutron
stars~\cite{Blaschke:2001,Sedrakian:2007}. Considering the spin
and isospin degrees of freedom, the NN Cooper pairs should have
rich inner structure and hence different phase diagrams.

Recently, two research directions in the study of nuclear matter
receive more attention. One is the BCS-BEC crossover
~\cite{Baldo:1995,Alm:1993,Lombardo:2001,Isayev:2004,Isayev:2006,Isayev:2008,Matsuo:2006,Hagino:2007,Margueron:2007}.
When nuclear density decreases, the weakly correlated NN BCS state
at high density may go over to the BEC superfluid of NN bound
state at lower density. Although the BCS and BEC limits are
physically quite different, the change from BCS to BEC was found
to be smooth~\cite{Leggett:1980,Nozieres:1985,deMelo:1993}. For
neutron-proton ($np$) pairs in $^3S_1-^3D_1$ channel, the chemical
potential changes sign at a critical density which can be regarded
as a criterion of the formation of BEC and finally approaches to a
half of the deuteron binding energy at low density
limit~\cite{Baldo:1995,Lombardo:2001}. Recently, the possible
BCS-BEC crossover of neutron-neutron ($nn$) pairs in $^1S_0$
channel was studied by examining the spatial structure of two
correlated neutrons~\cite{Matsuo:2006,Margueron:2007} and the
density and spin correlation functions~\cite{Isayev:2008}. It was
found that a di-neutron BEC state can be formed in symmetric
nuclear matter at very low neutron density. However, when the
degree of isospin asymmetry is high, there is no such a BEC state
in the whole density region.

The other direction is the possible phase transition induced by
the mismatch between neutron and proton Fermi
surfaces~\cite{Sedrakian:2000,Akhiezer:2001,Sedrakian:2001,Muther:2003,Alford:2005,Jin:2006}.
When the isospin asymmetry becomes sufficiently high, namely when
the mismatch is comparable with the $np$ pairing gap, the pairing
will be suppressed. A phase transition from the BCS state to
normal nuclear fluid is expected at a critical isospin asymmetry.
However, for an asymmetric system, besides the BCS state, some
other superfluid states are suggested in condensed matter and
nuclear matter, such as the Sarma phase~\cite{Sarma:1963} or
breached pairing phase~\cite{Liu:2003} where the superfluid
component is breached by the normal component in momentum space,
the Fulde-Ferrel-Larkin-Ovchinnikov (FFLO) phase~\cite{Fulde:1964}
where a Cooper pair has a total momentum and the translational
symmetry is spontaneously broken, the deformed Fermi surface
phase~\cite{Muther:2002,Sedrakian:2005} where the Fermi surfaces
of the two species are deformed into ellipsoidal shape and cross
to each other so that Cooper pairs with zero total momentum could
form at the cross node, and the phase separation (PS) in real
space~\cite{Bedaque:2003,Caldas:2004} where the normal and
superfluid components are inhomogeneously mixed.

Since nucleons could form Cooper pairs in both isospin singlet and
triplet channels, it is natural to ask a question of how the
competition between the isospin $I=0$ pairing, namely the $np$
pairing, and $I=1$ pairing, namely the $nn$ and $pp$ pairing,
affects the BCS-BEC crossover and thermodynamics of nuclear
superfluid. To have an insight into this question, we perform in
this paper a mean field analysis for the asymmetric nuclear matter
with the above two kinds of pairings. The mean field approximation
is an effective and successful treatment at low temperature or for
weakly correlated systems. At high temperature or for strongly
coupled systems, pair fluctuations may become
significant~\cite{Nozieres:1985}. We will restrict our study on
the BCS-BEC crossover at zero temperature and the thermodynamics
at high density. To simplify numerical calculations, We use a
density dependent contact potential to describe the NN
interaction.

The paper is organized as follows. We present the mean field
formalism for a general asymmetric nuclear matter with isospin
$I=0$ and $I=1$ pairings in Section \ref{2}. The possibility of
BCS-BEC crossover for the two channels is investigated in Section
\ref{3}, and the significant change in the phase structure of
weakly interacting nuclear superfluid at finite temperature is
discussed in Section \ref{4}. We summarize in section \ref{5}.

%%%%%%%%%%%%%%%%%%%%%%%%%%%%%%%%%%%%%%%%%%%%%%%%%%%%%%%%%%%%%%%%%%%%%%%
\section {Mean Field Formalism}\label{2}
\begin{figure}[htb]
\begin{center}
\includegraphics[width=7.5cm]{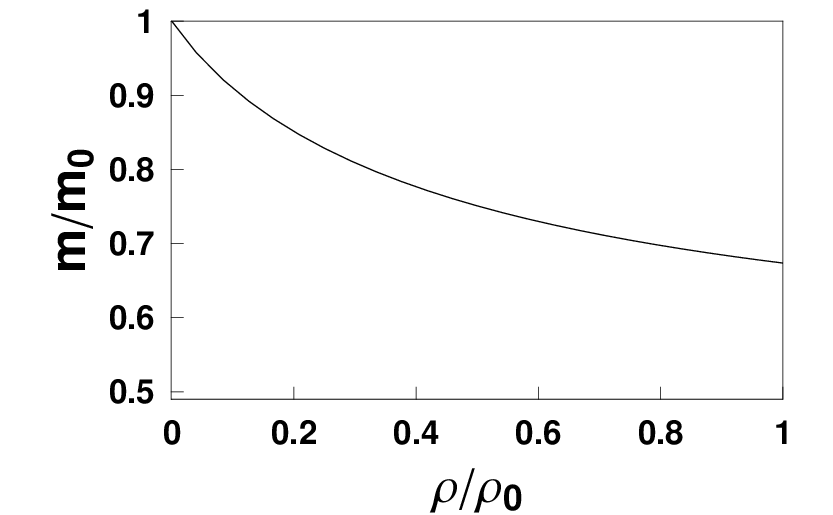}
\caption{The effective nucleon mass as a function of nuclear
density, calculated from the Gogny force D1. $m_0$ and $\r_0$ are,
respectively, nucleon mass in vacuum and normal nuclear density. }
\label{fig1}
\end{center}
\end{figure}

%%%%%%%%%%%%%%%%%%%%%%%%%%%%%%%%%%%%%%%%%%%%%%%%%%%%%%%%%%%%%%%%%%%%%%%
We adopt the density dependent contact interaction (DDCI)
developed by Gorrido et. al.~\cite{Garrido:1999} to model the NN
potential, due to its simplicity and validity in pairing problem.
For the purpose of getting qualitative conclusions which should
not be sensitive to the details of the interacting dynamics, the
DDCI potential is acceptable. The potential is of the form
\begin{equation}
\label{paris1}
V(\bx,\bx')=v\ls1-\w\lb\r((\bx+\bx')/2))\over\r_0\rb^{\g}\rs\d(\bx-\bx'),
\end{equation}
where $v, \w$ and $\g$ are three adjustable parameters,
$\r(\bx)=\r_n(\bx)+\r_p(\bx)$ is the nuclear density. Taking
suitable values of the parameters, one can
reproduce~\cite{Garrido:1999} the pairing gap $\D(k_F)$ as a
function of the Fermi momentum $k_F=(3\p^2\r/2)^{1/3}$ in the
channels $L=0,I=1,I_z=\pm1,S=0$ and $L=0,I=0,S=1,S_z=0$. We will
choose in the following numerical calculations the
parameters~\cite{Garrido:1999} $\w=0.45, \g=0.47, v=-481\ \rm{MeV
fm^3}$ in the $I=1$ channel and $\w=0, v=-530\ \rm{MeV fm^3}$ in
the $I=0$ channel and the energy cutoff $\e_c=60\ \rm{MeV}$ in
both channels to regularize the integration. With these parameters
one must use a density-dependent effective nucleon mass
$m(\r)$~\cite{Garrido:1999}, corresponding to the Gogny
interaction~\cite{Berger:1991},
\begin{eqnarray}
\label{m}
{m_0\over
m(\r)}&=&1+\frac{m_0}{2}\frac{k_F}{\sqrt{\p}}\sum_{c=1}^2[W_c+2(B_c-H_c)-4M_c]\non
&&\times\
\m_c^3e^{-x_c}\ls\frac{\cosh{x_c}}{x_c}-\frac{\sinh{x_c}}{x_c^2}\rs
\end{eqnarray}
with $x_c=k_F^2\m_c^2/2$, where $m_0=939\ \rm{MeV}$ is the nucleon
mass in vacuum, and $\m_c, W_c, B_c, H_c, M_c$ are parameters by
fitting the Gogny force D1~\cite{Ring:1980,Decharge:1980}, their
values are listed in Table~\ref{tab1}. As shown in \fig{fig1}, the
medium effect suppresses the nucleon mass.
%%%%%%%%%%%%%%%%%%%%%%%%%%%%%%%%%%%%%%%%%%%%%%%%%%%%%%%%%%%%%%%%%%%%%%%%
\begin{table}[htb]
\caption{Parameters in the effective nucleon mass (\ref{m}) by
fitting the Gogny interaction D1~\cite{Ring:1980,Decharge:1980}.}
\begin{tabular}{cccccc} \hline\hline
c\ & $\m_c$[fm]\ & $W_c$[MeV]\ & $B_c$[MeV]\ & $H_c$[MeV]\ & $M_c$[MeV]\\
\hline 1 & 0.7 & -402.4 & -100.0 & -496.2 & -23.56\\
2 & 1.2 & -21.30 & -11.77 & 37.27 & -68.81\\
\hline\hline
\end{tabular}\label{tab1}
\end{table}
%%%%%%%%%%%%%%%%%%%%%%%%%%%%%%%%%%%%%%%%%%%%%%%%%%%%%%%%%%%%%%%%%%%%%%%%
%%%%%%%%%%%%%%%%%%%%%%%%%%%%%%%%%%%%%%%%%%%%%%%%%%%%%%%%%%%%%%%%%%%%%%%
%%%%%%%%%%%%%%%%%%%%%%%%%%%%%%%%%%%%%%%%%%%%%%%%%%%%%%%%%%%%%%%%%%%%%%%

For a uniform nuclear system, the nuclear density $\r$ is
independent of $\bx$, and the DDCI potential is simplified to the
form of
\begin{equation}
V_I(\bx-\bx')=g_I\d(\bx-\bx')
\end{equation}
with the effective coupling constant
\begin{equation}
g_I=v_I[1-\w_I(\r/\r_0)^{\g_I}],
\end{equation}
where $I=0,1$ denote the isospin of pairs.

The uniform nuclear system with the two-body interacting potential
$V_I(\bx-\bx')$ and chemical potentials $\m_n$ and $\m_p$ for
neutrons and protons is controlled by the Lagrangian density
\begin{widetext}
\begin{eqnarray}
\label{lagrangian} \hat\cl&=&\sum_{\s=\ua,\da}\ls \hat
p_\s^\dag(\bx)\left(-\frac{\pt}{\pt\t}+\frac{\nabla^2}{2m}+\m_p\right)\hat
p_\s(\bx)+\hat
n_\s^\dag(\bx)\left(-\frac{\pt}{\pt\t}+\frac{\nabla^2}{2m}+\m_n\right)
\hat n_\s(\bx)\rs\non &&-\int d^3\bx'V_1(\bx-\bx')\ls \hat
n^{\dag}_\ua(\bx) \hat n^{\dag}_\da(\bx')\hat n_\da(\bx')\hat
n_\ua(\bx)+\hat p^{\dag}_\ua(\bx)\hat p^{\dag}_\da(\bx')\hat
p_\da(\bx')\hat p_\ua(\bx)\rs\non &&-\frac{1}{2}\int
d^3\bx'V_0(\bx-\bx')\ls \hat n^{\dag}_\ua(\bx)\hat
p^{\dag}_\da(\bx')-\hat p^{\dag}_\ua(\bx)\hat
n^{\dag}_\da(\bx')\rs \Big[\hat p_\da(\bx')\hat n_\ua(\bx)-\hat
n_\da(\bx')\hat p_\ua(\bx)\Big]\non &=&\sum_{\s=\ua,\da}\ls \hat
p_\s^\dag(\bx)\left(-\frac{\pt}{\pt\t}+\frac{\nabla^2}{2m}+\m_p\right)\hat
p_\s(\bx)+\hat
n_\s^\dag(\bx)\left(-\frac{\pt}{\pt\t}+\frac{\nabla^2}{2m}+\m_n\right)
\hat n_\s(\bx)\rs\non &&-g_1\left(\hat n^{\dag}_\ua(\bx) \hat
n^{\dag}_\da(\bx) \hat n_\da(\bx) \hat n_\ua(\bx)+\hat
p^{\dag}_\ua(\bx) \hat p^{\dag}_\da(\bx) \hat p_\da(\bx) \hat
p_\ua(\bx)\right)\non &&-\frac{1}{2}g_0\left(\hat
n^{\dag}_\ua(\bx) \hat p^{\dag}_\da(\bx)-\hat p^{\dag}_\ua(\bx)
\hat n^{\dag}_\da(\bx)\right)\Big(\hat p_\da(\bx) \hat
n_\ua(\bx)-\hat n_\da(\bx) \hat p_\ua(\bx)\Big),
\end{eqnarray}
\end{widetext}
where $\hat n_\s$ and $\hat p_\s$ are neutron and proton field
operators with spin $\s$. We introduce the Cooper pair operators
$\hat\D_{np}(\bx)=-(g_0/2)(\hat p_\da(\bx) \hat n_\ua(\bx)-\hat
n_\da(\bx) \hat p_\ua(\bx))=\hat\D_{np}^*(\bx),\
\hat\D_{nn}(\bx)=-g_1\hat n_\da(\bx) \hat
n_\ua(\bx)=\hat\D_{nn}^*(\bx)$ and $\hat\D_{pp}(\bx)=-g_1\hat
p_\da(\bx) \hat p_\ua(\bx)=\hat\D_{pp}^*(\bx)$, corresponding to
the channels $L=0,I=0,S=1,S_z=0$ and $L=0,I=1,I_z=\pm,S=0$, and
keep only their condensates $\langle \hat\D_{np}(\bx)\rangle,\
\langle \hat\D_{nn}(\bx)\rangle$ and $\langle
\hat\D_{pp}(\bx)\rangle$ in the Lagrangian in mean field
approximation. To incorporate the FFLO state into our study, we
assume the following forms of the condensates,
$\langle\hat\D_{np}(\bx)\rangle=\D_{np}
e^{i(\bq_n+\bq_p)\cdot\bx},\
\langle\hat\D_{nn}(\bx)\rangle=\D_{nn} e^{2i\bq_n\cdot\bx}$ and
$\langle\hat\D_{pp}(\bx)\rangle=\D_{pp} e^{2i\bq_p\cdot\bx}$,
where $\D_{np},\ \D_{nn}$ and $\D_{pp}$ are constants and can be
assumed to be real numbers, and $\bq_n$ and $\bq_p$ are the pair
momenta. Obviously, the translational symmetry and rotational
symmetry in the FFLO state are spontaneously broken. Note that,
for the sake of simplicity, the FFLO state we considered here is
its simplest pattern, namely the single plane wave FFLO state or
the so-called FF state.

The partition function $Z$ which is the key quantity of a
thermodynamic system can be calculated by path integral,
\begin{equation}
\label{Z} Z=\P_{\s}\int [d\hat n_\s] [d\hat p_\s] [d\hat
n_\s^\dag] [d\hat p_\s^\dag]\exp{\lb\int_0^\b d\t \int
d^3\bx\;\hat\cl\rb}.
\end{equation}
By performing a gauge transformation for the nucleon fields
$\tilde n_\s=e^{-i\bq_n\cdot\bx}\hat n_\s$ and $\tilde
p_\s=e^{-i\bq_p\cdot\bx}\hat p_\s$ which keeps the functional
measure in \eq{Z} invariant, the path integral over $\tilde n_\s$
and $\tilde p_\s$ can be easily done and we obtain the mean field
thermodynamic potential
\begin{eqnarray}
\label{omega1} \O&=&-\frac{T}{V}\ln
Z\non&=&-\frac{2\D_{np}^2}{g_0}-\frac{\D_{nn}^2+\D_{pp}^2}{g_1}\non&&-T\sum_{\n}\int\frac{d^3\bk}{(2\p)^3}\Tr\ln
G^{-1}(i\o_\n,\bk),
\end{eqnarray}
where $\o_\n=(2\n+1)\p T$ with $\n\in\mathbb{Z}$ is the fermion
frequency, and $G$ is the Nambu-Gorkov propagator,
\begin{equation}
G^{-1}=\left(\begin{array}{cccc}i\omega_\n-\epsilon_n^+ & 0 & \Delta_{np} & \Delta_{nn}\\
0 & i\omega_\n-\epsilon_p^+ & \Delta_{pp} & -\Delta_{np}\\
\Delta_{np} & \Delta_{pp} & i\omega_\n+\epsilon_p^- & 0\\
\Delta_{nn} & -\Delta_{np} & 0 &i\omega_\n+\epsilon_n^-
\end{array}\right)
\end{equation}
with the definition $\epsilon_i^\pm=({\bf k}\pm{\bf
q}_i)^2/(2m)-\mu_i$ for $i=n,p$. It is worthy noting that, if the
relative momentum $\bq_n-\bq_p$ is large enough, the uniform
superfluid would be unstable due to the stratification of the
superfluid components characterized by $\D_{nn}$ and $\D_{pp}$, in
analogous to the multi-component BEC in condensed matter physics.
To avoid such a dynamic instability~\cite{Khalatnikov:1973}, we
choose $\bq_n=\bq_p=\bq$. After computing the frequency summation
and trace in (\ref{omega1}), the thermodynamic potential can be
expressed in terms of quasi-particles,
\begin{eqnarray}
\label{omega} \Omega &=&
-\frac{2\D_{np}^2}{g_0}-\frac{\D_{nn}^2+\D_{pp}^2}{g_1}+\int\frac{d^3\bk}{(2\p)^3}\Bigg[\epsilon_n^-+\epsilon_p^-
\non&&-\sum_{i,j=\pm}\left({E_j^i\over
2}+T\ln\left(1+e^{-E_j^i/T}\right)\right)\Bigg],
\end{eqnarray}
where $E_\pm^\mp$ are the quasi-particle energies
\begin{equation}
E_\pm^\mp=\sqrt{\epsilon_+^2+\delta\epsilon^2\pm\sqrt{\epsilon_-^4+\epsilon_\Delta^4}}\mp\delta\epsilon
\end{equation}
with $\epsilon_\pm, \delta\epsilon$ and $\epsilon_\Delta$ defined as
\begin{eqnarray}
&& 2\epsilon_\pm^2 =
\epsilon_n^+\epsilon_n^-+\Delta_{nn}^2+\Delta_{np}^2\pm\left(\epsilon_p^+\epsilon_p^-+\Delta_{pp}^2+\Delta_{np}^2\right)
,\nonumber\\
&& \delta\epsilon =
(\epsilon_n^+-\epsilon_n^-)/2=(\epsilon_p^+-\epsilon_p^-)/2,\\
&& \epsilon_\Delta^4 =
\Delta_{np}^2[(\epsilon_n^+-\epsilon_p^+)(\epsilon_n^--\epsilon_p^-)+(\Delta_{pp}-\Delta_{nn})^2].\nonumber
\end{eqnarray}

From the thermodynamic potential, we derive the neutron and proton
number densities,
\begin{equation}
\label{densityeq1} \r_n=-{\partial\Omega\over
\partial\m_n},\ \ \ \ \r_p=-{\partial\Omega\over \partial\m_p},
\end{equation}
or
\begin{equation}
\label{densityeq2} \r=-{\partial\Omega\over \partial\m},\ \ \ \
\d\r= -{\partial\Omega\over \partial\d\m},
\end{equation}
where $\m=(\m_n+\m_p)/2$ is the averaged nucleon chemical
potential, $\d\m=(\m_n-\m_p)/2$ the mismatch between $\m_n$ and
$\m_p$, and $\d\r=\r_n-\r_p$ the associated difference in number
densities. We will focus on neutron rich nuclear matter with
$\d\m>0$ and $\d\r>0$, since it is coincident with the environment
of neutron stars and some heavy nuclei. For the sake of
convenience, we define the relative density asymmetry
$\a=\d\r/\r$.

The condensates and the FFLO momentum are determined by the gap
equations,
\begin{equation}
\label{gapeq} {\partial\Omega\over \partial\Delta_{np}}=0,\ \
{\partial\Omega\over \partial\Delta_{nn}}=0,\ \
{\partial\Omega\over
\partial\Delta_{pp}}=0,\ \
{\partial\Omega\over \partial{\bf q}}=0.
\end{equation}
With the above coupled number equations (\ref{densityeq2}) and gap
equations (\ref{gapeq}), we can solve the condensates
$\Delta_{np},\ \Delta_{nn},\ \Delta_{pp}$, FFLO momentum ${\bf q}$
and chemical potentials $\mu,\ \delta\mu$ as functions of
temperature $T$ and densities $\r,\ \d\r$. The ground state of the
system is specified by the solution corresponding to the global
minimum of the free energy which is related to the thermodynamic
potential by a Legendre transformation,
$\cf=\O+\m_n\r_n+\m_p\r_p=\O+\m\r+\d\m\d\r$.

%%%%%%%%%%%%%%%%%%%%%%%%%%%%%%%%%%%%%%%%%%%%%%%%%%%%%%%%%%%%%%%%%%%%%%%
\section {BCS-BEC Crossover at Zero Temperature}\label{3}
%%%%%%%%%%%%%%%%%%%%%%%%%%%%%%%%%%%%%%%%%%%%%%%%%%%%%%%%%%%%%%%%%%%%%%%
The mean field approximation has been successfully used to study
the BCS-BEC crossover in nuclear matter with isospin
$I=0$~\cite{Baldo:1995,Alm:1993,Lombardo:2001,Isayev:2004,Isayev:2006}
or $I=1$~\cite{Isayev:2008,Matsuo:2006,Hagino:2007,Margueron:2007}
pairings. We extend in this section the study to a more general
case with both $I=0$ and $I=1$ pairings. To guarantee the validity
of mean field treatment, we focus on the superfluid at zero
temperature. Since the FFLO state is unstable in strong coupling
region~\cite{Sheehy:2006}, we will not take into account the FFLO
state in the study of BCS-BEC crossover. The temperature effect
and the FFLO state will be considered in Section \ref{4}.

To examine how the NN correlation changes with nuclear density, we
define the normalized Cooper pair wave function as
\begin{eqnarray}
\label{wavefunc} \j_{ij}(\br)&=&C\lan BCS|\hat
a^\dag_{i\ua}(\bx)\hat a^\dag_{j\da}(\bx+\br)|BCS\ran\non
&=&C'\idp{3}{\bk}\j_{ij}(\bk)e^{i\bk\cdot\br}
\end{eqnarray}
with $i,j=n,p$, where $\hat a_{i\s}^\dag$ is the nucleon creation
operator, and $\j_{ij}(\bk)$ the wave function in momentum space
or the anomalous density
\begin{equation}
\label{anomalousdensity} \j_{ij}(\bk)=\lan BCS|\hat
a^\dag_{i\ua}(\bk)\hat a^\dag_{j\da}(-\bk)|BCS\ran.
\end{equation}
The density distribution functions $n_i(\bk)$ is defined through
\begin{equation}
\label{normaldensity} n_{i}(\bk)=\frac{1}{2}\sum_\s\lan BCS|\hat
a^\dag_{i\s}(\bk)\hat a_{i\s}(\bk)|BCS\ran.
\end{equation}

In order to describe the BCS-BEC crossover quantitatively, it is
convenient to introduce the following characteristic
quantities\cite{Matsuo:2006,Margueron:2007}:\\
1) The distribution function $r^2|\j_{ij}(\br)|^2$ which is the
probability to find a pair of nucleons $i$ and $j$ with distance
$r$ in between. For the BCS-BEC crossover induced by changing
coupling constant at fixed density such as in cold atom gas, the
probability would distribute in a wide space region in the weakly
coupled BCS state but peak sharply at a small $r$ in the strongly
coupled BEC state. However, for nuclear superfluid, the BCS-BEC
crossover is induced by changing the nuclear density. The density
dependence of any physical quantity of the system is reflected in
two aspects: the explicit or direct $\r$ dependence through the
coupling between the gap and number equations and the indirect
$\r$ dependence through the effective coupling constant $g_I(\r)$
and in-medium nucleon mass $m(\r)$. When the density drops down,
the indirect $\r$ dependence is the driving force for the
crossover from BCS to BEC, but the explicit $\r$ dependence makes
the system dilute and then slows down or even rejects the
crossover. The balance between these two opposite aspects controls
the BCS-BEC crossover and will probably change the monotonous
shrinking behavior of
the probability.\\
2) The root-mean-square radius of the Cooper pair
$\x_{ij}=\sqrt{\lan r^2\ran_{ij}}$ with $\lan r^2\ran_{ij}=\int
d^3{\bf r}\ r^2|\j_{ij}(\br)|^2$ which characterizes the size of
the Cooper pair. When the density is fixed, $\x_{ij}$ is expected
to be large in the weakly coupled BCS region and small in the
strongly coupled BEC region. However, for nuclear superfluid with
decreasing nuclear density from BCS to BEC, $\x_{ij}$ itself can
no longer describe the BCS-BEC crossover. In this case, the BCS
and BEC states are defined in the sense of overlapping degree of
pair wave functions. At high density the size of a pair can be
small but the pairs may overlap strongly, while at low density the
size of a pair is probably large but the overlapping of pairs
becomes weak. To represent the density dependence of the
overlapping, we can compare $\x_{ij}$ with the averaged distance
$d_{ij}=\r_{ij}^{-1/3}$ between any two nucleons with
$\r_{np}=\r/2,\ \r_{nn}=\r_n$ and $\r_{pp}=\r_p$. For $\x_{ij}\gg
d_{ij}$ the overlapping is strong, and the $ij-$pair can be
interpreted as an extended BCS pair, while for $\x_{ij}\ll d_{ij}$
the overlapping becomes weak, and the pair should be considered as
a compact BEC pair, \ie, a boson-like bound state. \\
3) The probability $P(d_{ij})=4\pi\int_0^{d_{ij}} dr\ r^2
|\j_{ij}(\br)|^2$ of finding an $ij-$Cooper pair within the
averaged distance $d_{ij}$ between the nucleons $i$ and $j$. Note
that the probability is density dependent and can be used to
describe the overlapping degree of pairs. It should be very close
to 1 in the strongly coupled BEC state and clearly less than 1
when the
superfluid is in the BCS state.\\
4) The s-wave scattering length $a_{ij}$ which relates the
coupling constant $g_I$ to the low energy limit of the T-matrix
for nucleons $i$ and $j$ in vacuum,
\begin{equation}
\label{aij} \frac{m}{4\p a_{ij}}=\frac{1}{g_{ij}}+\int{d^3{\bf
k}\over (2\pi)^3}\frac{1}{2\e_\bk}
\end{equation}
with $\e_\bk=\bk^2/(2m)$ and $g_{np}=g_0, g_{nn}=g_{pp}=g_1$. In
the BCS region $a_{ij}$ is negative, representing the attractive
interaction between nucleons. However, in the BEC region it should
be positive to preserve the stability of the two-body bound state.
Therefore, the change of $a_{ij}$ from negative to positive value
can be considered as a signature of the BCS-BEC crossover. \\
5) The scaled condensate $\D_{ij}/\e^{ij}_F$ with the Fermi energy
defined as $\e^{ij}_F=(k^{ij}_F)^2/(2m)$, where $k^{np}_F=k_F,\
k^{nn}_F=k_{nF}$ and $k^{pp}_F=k_{pF}$ are the corresponding Fermi
momenta. In the case with only $I=0$ pairs, the mean field gap
equation and number equation satisfy the so-called universality:
The scaled quantities like $\D/\e_F$ and $\m/\e_F$ are only
functions of the effective coupling $1/(k_F
a)$~\cite{Leggett:1980}. In the strongly coupled BEC region, one
should expect a large $\D/\e_F$, but in the weakly interacting BCS
state it will be less than $1$. While the universality is
explicitly broken in our general case with both $I=0$ and $I=1$
pairings, we still take the scaled quantities to describe the BCS-BEC crossover. \\
6) The effective chemical potential $\m_{ij}/\e^{ij}_F$ with
$\m_{np}=\m, \m_{nn}=\m_n$ and $\m_{pp}=\m_p$. Due to the
definition, $\m_{ij}/\e^{ij}_F$ is exactly equal to $1$ in the BCS
limit and the order of 1 in the weakly coupled BCS region. In the
case of only one kind of pairs, like symmetric nuclear matter and
neutron matter discussed below, $\m/\e_F$ is negative in the BEC
region, the absolute value $|\m/\e_F|$ becomes larger and larger
as the system goes deeper and deeper into the BEC region, and
$2\m$ can be viewed as the binding energy of the bound state in
the BEC limit when the system approaches to vanishing density.
That is the reason why the change of the sign of $\m$ is normally
considered as an unambiguous criterion of the formation of BEC, at
least at mean field level. However, in general asymmetric nuclear
matter with $np,\ nn$ and $pp$ pairs, the correlations among
different pairs will qualitatively change this conclusion, see the
detailed discussion in the following.

%%%%%%%%%%%%%%%%%%%%%%%%%%%%%%%%%%%%%%%%%%%%%%%%%%%%%%%%%%%%%%%%%%%%%%%
\subsection {Symmetric Nuclear Matter: $\a=0$}\label{3.1}
%%%%%%%%%%%%%%%%%%%%%%%%%%%%%%%%%%%%%%%%%%%%%%%%%%%%%%%%%%%%%%%%%%%%%%%
Let us first consider the symmetric nuclear matter with $\d\r=0,\
\d\m=0$ and $\D_{nn}=\D_{pp}$. The number and gap equations in
this case are largely simplified as
\begin{eqnarray}
\label{a0} \r_n&=&\r_p=2\idp{3}{\bk}n(\bk),\non n(\bk) &=&
n_n(\bk)=n_p(\bk)=\frac{1}{2}\lb1-\frac{\e_\bk}{E_\bk}\rb,\non
\D_{np}&=&-g_0\D_{np}\idp{3}{\bk}\frac{1}{2E_\bk},\non
\D_{nn}&=&\D_{pp}=-g_1\D_{nn}\idp{3}{\bk}\frac{1}{2E_\bk},
\end{eqnarray}
where $\e_\bk=\bk^2/(2m)-\m$ is the particle dispersion relation
and $E_\bk=\sqrt{\e_\bk^2+\D_{np}^2+\D_{nn}^2}$ the quasi-particle
dispersion. The gap equations have two types of nontrivial
solutions $\D_{np}\neq 0, \D_{nn}=\D_{pp}=0$ and $\D_{np}=0,
\D_{nn}=\D_{pp}\neq 0$. Considering $|g_0|>|g_1|$ at any density,
only the solution with $\D_{np}\neq 0$ corresponds to the ground
state. This can also be verified by comparing the free energies
for the two solutions. Since there is no experimental evidence for
$np$ pairing at finite nuclear density, it is necessary to note
that the condition $|g_0|>|g_1|$ is in principle an assumption in
our treatment. From a more self-consistent Greens function
approach~\cite{muther:2006}, the tensor correlations in the $I=0$
channel may yield a kind of pair correlation which is different
from the one observed in solving gap equations.

To simplify the notification, we write in short $\e_F=\e_F^{np}$
and $d=d_{np}$ in the following. Solving the coupled number and
gap equations for the $np$ channel, we obtain the scaled gap
$\D_{np}(\r)/\e_F$ and chemical potential $\m(\r)/\e_F$ as
functions of nuclear density.

Defining the anomalous density $\j_{np}(\bk) = \D_{np}/(2E_\bk)$
and substituting it into the number and gap equations give the
Schr\"odinger-like equation
\begin{equation}
\label{schronp}
\frac{\bk^2}{m}\j_{np}(\bk)+(1-2n_\bk)g_0\idp{3}{\bk'}
\j_{np}(\bk')=2\m\j_{np}(\bk)
\end{equation}
for the anomalous density. In the limit of vanishing density,
$n_\bk\ra 0$, the equation goes over into the Schr\"odinger
equation for the $np$ bound state, and the chemical potential
$2\m$ plays the role of binding energy. Since such a bound state
should be a boson, one expects that, at sufficiently low density
and low temperature the symmetric nuclear matter is in the BEC
phase.

%%%%%%%%%%%%%%%%%%%%%%%%%%%%%%%%%%%%%%%%%%%%%%%%%%%%%%%%%%%%%%%%%%%%%%%
\begin{figure}[!htb]
\begin{center}
\includegraphics[width=8cm]{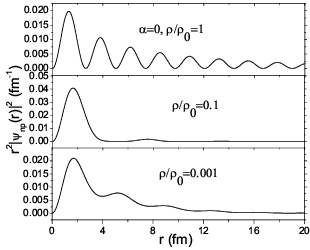}
\caption{The probability $r^2|\j_{np}(r)|^2$ as a function of the
relative distance $r$ between the paired neutron and proton at
different nuclear density $\r$ in symmetric nuclear matter. }
\label{fig2}
\end{center}
\end{figure}
%%%%%%%%%%%%%%%%%%%%%%%%%%%%%%%%%%%%%%%%%%%%%%%%%%%%%%%%%%%%%%%%%%%%%%%

\begin{figure}[!htb]
\begin{center}
\includegraphics[width=7cm]{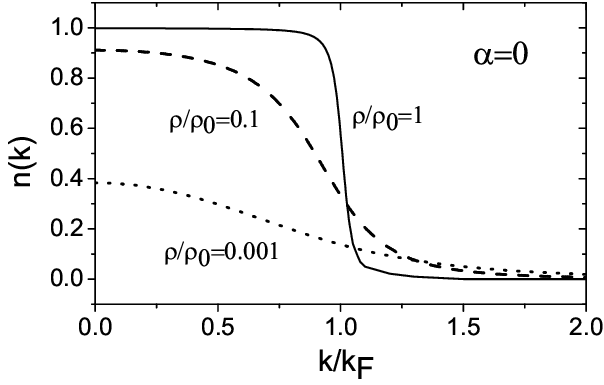}
\caption{ The nuclear density distribution $n(k)$ as a function of
momentum at fixed nuclear density $\r$ in symmetric nuclear
matter. $k_F$ is the Fermi momentum.}
\label{fig3}
\end{center}
\end{figure}

Making the Fourier transformation of the anomalous density
$\psi_{np}(\bk)$ which is determined by solving the gap and number
equations, we obtain the wave function $\psi_{np}(\br)$ in
coordinate space. The behavior of the probability distribution
$r^2|\j_{np}(\br)|^2$ as a function of the relative distance $r$
between the pair partners is shown in \fig{fig2}. The spatial
extension and profile of the probability depend strongly on the
density. Near the normal density, the probability function is
spatially extended and behaves like the well-known BCS
expression~\cite{Bardeen:1957} $\j_{np}(r)\sim
K_0(r/\p\x_{np})\sin(k_F r)/(k_Fr)$. The oscillation induced by
the sharp Fermi surface at high density is well-known and called
Friedel oscillation~\cite{Friedel:1954} which is widely discussed
in nuclear matter and recently extended to quark
matter~\cite{friedel}. The Fermi surface is strictly defined only
in normal state without any condensate. In the BCS region at high
density, the condensate is small and there is still an approximate
Fermi surface. However, in the BEC region at low density, the
Fermi surface is already very low and further deformed by the
large condensate. The number density as a function of momentum is
plotted in \fig{fig3}. It is clear that the Fermi surface is
well-defined only at high density. This is the reason why the
Friedel oscillation almost disappears at low density.
%%%%%%%%%%%%%%%%%%%%%%%%%%%%%%%%%%%%%%%%%%%%%%%%%%%%%%%%%%%%%%%%%%%%%%%
%%%%%%%%%%%%%%%%%%%%%%%%%%%%%%%%%%%%%%%%%%%%%%%%%%%%%%%%%%%%%%%%%%%%%%%

It is necessary to note that the probability $r^2|\j_{np}(r)|^2$
does not shrink monotonously with decreasing number density, see
\fig{fig2}. For a very dilute system, the two partners of a pair
can have a large distance between them, the pair wave function
will be spatially extended, and the probability can then
distribute in a wide space region. While the strong Friedel
oscillation at high density and its weakening at low density can
be considered to distinguish the BCS state from the BEC state, the
space extension of the probability itself can not be used to
characterize the BCS-BEC crossover in nuclear superfluid.

\begin{figure}[!htb]
\begin{center}
\includegraphics[width=7cm]{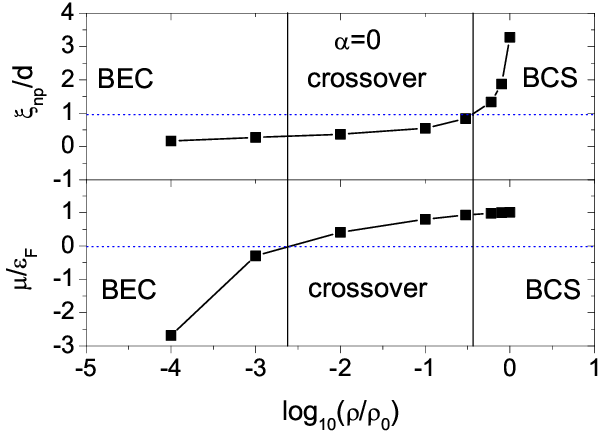}
\caption{ The scaled root-mean-square radius $\x_{np}/d$ and
scaled chemical potential $\m/\e_F$ as functions of nuclear
density in symmetric nuclear matter. $\e_F$ is the nucleon Fermi
energy and $d$ the averaged distance between two nucleons. The
right and left vertical lines which separate the BCS, BEC and
crossover regions are, respectively, determined by the conditions
$\x_{np}/d=1$ and $\m/\e_F=0$. }
\label{fig4}
\end{center}
\end{figure}

To clearly describe the BCS-BEC crossover induced by the change in
nuclear density, we now calculate the scaled root-mean-square
radius $\x_{np}/d$, the scaled chemical potential $\m/\e_F$, the
scaled gap parameter $\D_{np}/\e_F$, the effective scattering
length $1/(k_F a_{np})$ and the probability $P_{np}(d)$.
$\x_{np}/d$ and $\m/\e_F$ are presented in \fig{fig4} as functions
of nuclear density. While $\x_{np}$ itself is not a monotonous
function of $\r$, the scaled one goes up monotonously with
increasing density. The right vertical line at $\r/\r_0 \sim 0.4$
indicates the position of $\x_{np}/d=1$ which can be used to
separate the BCS region with sharply increasing $\x_{np}/d$ from
the region with slightly changing and small $\x_{np}/d$. The
scaled chemical potential is $1$ in the BCS limit, then drops down
with decreasing density, and becomes negative at very small
density. The position where $\m/\e_F$ approaches to zero is
indicated by the left vertical line at $\r/\r_0\sim 0.003$ which
is, from the definition, considered to distinguish the BEC region
with negative $\m$ from the other region with positive $\m$.
Therefore, considering both $\x_{np}/d$ and $\m/\e_F$, the BCS and
BEC states are, respectively, located at $\r/\r_0 > 0.4$ and
$\r/\r_0 < 0.003$, and the crossover from BCS to BEC is in between
the two vertical lines.
%%%%%%%%%%%%%%%%%%%%%%%%%%%%%%%%%%%%%%%%%%%%%%%%%%%%%%%%%%%%%%%%%%%%%%%
%%%%%%%%%%%%%%%%%%%%%%%%%%%%%%%%%%%%%%%%%%%%%%%%%%%%%%%%%%%%%%%%%%%%%%%

\begin{figure}[!htb]
\begin{center}
\includegraphics[width=7cm]{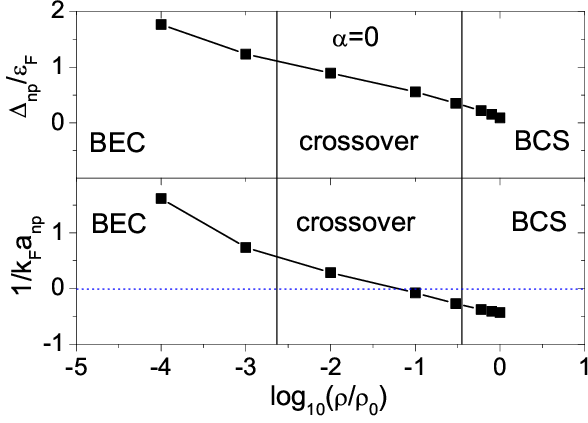}
\caption{ The scaled condensate $\D_{np}/\e_F$ and effective
scattering length $1/(k_Fa_{np})$ as functions of nuclear density
in symmetric nuclear matter. $a_{np}$ is the neutron-proton
scattering length. }
\label{fig5}
\end{center}
\end{figure}

In \fig{fig5} we show the scaled condensate $\D_{np}/\e_F$ and the
effective s-wave scattering length $1/(k_Fa_{np})$. Both are
monotonous functions of $\r$, while again the gap parameter
$\D_{np}$ itself does not behave monotonously. The scaled gap is
large at low density and small at high density and approaches to
zero in the BCS limit. The effective scattering length drops down
with increasing density and becomes negative at high density. The
density dependence of $\D_{np}/\e_F$ and $1/(k_F a_{np})$ agrees
well with the normal understanding of both BCS state with small
gap and negative scattering length and BEC state with large gap
and large and positive scattering length.

\begin{figure}[!htb]
\begin{center}
\includegraphics[width=7cm]{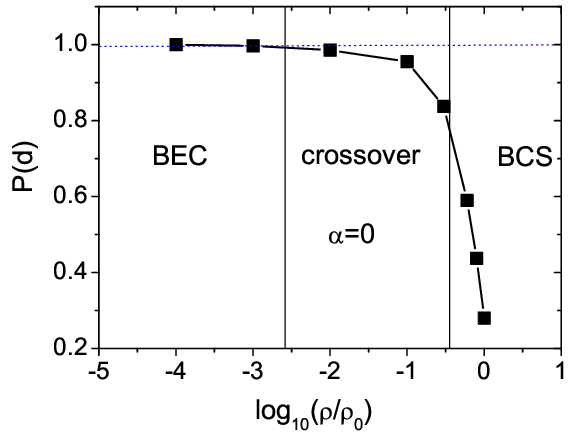}
\caption{ The probability $P_{np}(d)$ of finding a $np$-pair with
relative distance $r\leq d$ between the paired neutron and proton
as a function of nuclear density in symmetric nuclear matter. }
\label{fig6}
\end{center}
\end{figure}

The probability $P_{np}(d)$ of finding a pair with relative
distance $r\leq d$ between the paired neutron and proton is
plotted in \fig{fig6}. Again, while the probability $P_{np}(r)$ is
not a monotonous function of $\r$ for fixed $r$, $P_{np}(d)$
decreases monotonously with increasing density. It approaches to
$1$ at low density which means strong correlation of pairs in the
BEC state and becomes very small at high density which indicates a
weak correlation of pairs in the BCS state. The two vertical
straight lines in \fig{fig5} and \fig{fig6} which separate the
BCS, BEC and crossover regions are still characterized by the
scaled root-mean-square radius and scaled chemical potential. By
comparing the above three figures, we can see that $\x_{np}/d,\
\m/\e_F,\ \D_{np}/\e_F,\ 1/(k_Fa_{np})$ and $P_{np}(d)$ can
self-consistently describe the BCS-BEC crossover in symmetric
nuclear superfluid.
%%%%%%%%%%%%%%%%%%%%%%%%%%%%%%%%%%%%%%%%%%%%%%%%%%%%%%%%%%%%%%%%%%%%%%%
%%%%%%%%%%%%%%%%%%%%%%%%%%%%%%%%%%%%%%%%%%%%%%%%%%%%%%%%%%%%%%%%%%%%%%%
%%%%%%%%%%%%%%%%%%%%%%%%%%%%%%%%%%%%%%%%%%%%%%%%%%%%%%%%%%%%%%%%%%%%%%%
%%%%%%%%%%%%%%%%%%%%%%%%%%%%%%%%%%%%%%%%%%%%%%%%%%%%%%%%%%%%%%%%%%%%%%%

%%%%%%%%%%%%%%%%%%%%%%%%%%%%%%%%%%%%%%%%%%%%%%%%%%%%%%%%%%%%%%%%%%%%%%%
\subsection {Neutron Matter: $\a=1$}\label{3.2}
%%%%%%%%%%%%%%%%%%%%%%%%%%%%%%%%%%%%%%%%%%%%%%%%%%%%%%%%%%%%%%%%%%%%%%%
Neutron matter, as fully asymmetric nuclear matter, is important
for physics of neutron stars. In this case, there are $\a=1,
\r_n=\r,\ \r_p=0$ and $\D_{np}=\D_{pp}=0$. By solving the coupled
number equation and gap equation for $\r_{n}$ and $\D_{nn}$ which
are exactly the same as Eq.(\ref{a0}) when we replace $\e_{\bk}$
and $E_{\bk}$ by $\e_{n\bk}=\bk^2/(2m)-\m_n$ and
$E_{n\bk}=\sqrt{\e_{n\bk}^2+\D_{nn}^2}$, we obtain the neutron
chemical potential $\mu_n$ and $nn$ pair condensate $\D_{nn}$ as
functions of nuclear density. The anomalous density
$\j_{nn}(\bk)=\D_{nn}/(2E_{n\bk})$ satisfies the similar
Schr\"odinger-like equation (\ref{schronp}), and in the low
density limit $2\m_n$ plays the role of binding energy of the
possible di-neutron bound state.

\begin{figure}[!htb]
\begin{center}
\includegraphics[width=7cm]{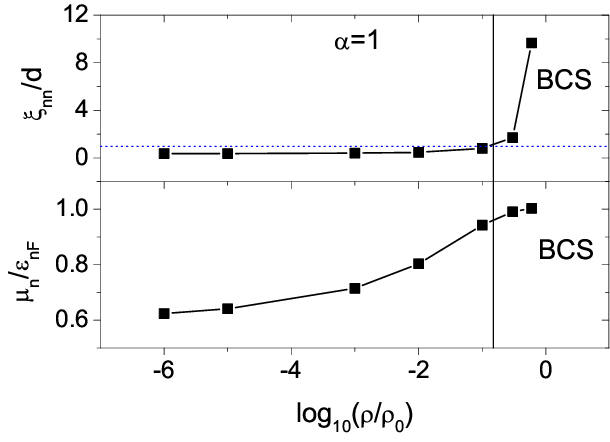}
\caption{ The scaled root-mean-square radius $\x_{nn}/d_{nn}$ and
neutron chemical potential $\m_n/\e_{nF}$ as functions of nuclear
density in neutron matter. The vertical line represents the BCS
boundary. $d_{nn}$ is the averaged distance between two neutrons
and $\e_{nF}$ the neutron Fermi energy. }
\label{fig7}
\end{center}
\end{figure}

\begin{figure}[!htb]
\begin{center}
\includegraphics[width=7cm]{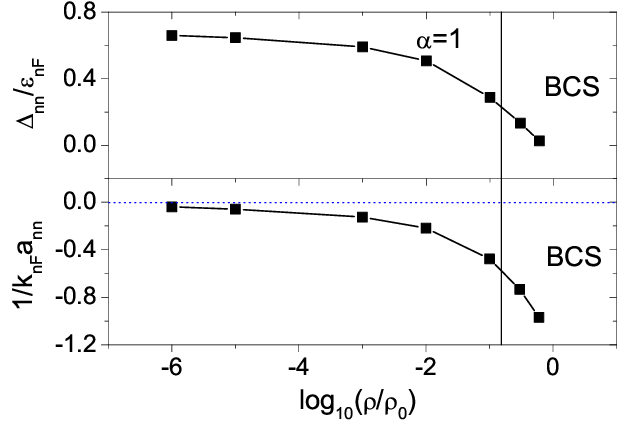}
\caption{ The scaled condensate $\D_{nn}/\e_F$ and effective
scattering length $1/(k_Fa_{nn})$ as functions of nuclear density
in neutron matter. $a_{nn}$ is the neutron-neutron scattering
length. }
\label{fig8}
\end{center}
\end{figure}

In comparison with the symmetric nuclear matter, the probability
$r^2|\j_{nn}(\br)|^2$ and the density distribution $n_n(\bk)$ at
different total number density are very similar to
$r^2|\j_{np}(\br)|^2$ and $n(\bk)$ shown in \fig{fig2} and
\fig{fig3}, but the scaled root-mean-square radius
$\xi_{nn}/d_{nn}$, chemical potential $\m_n/\e_{nF}$, gap
parameter $\D_{nn}/\e_{nF}$ and the effective scattering length
$1/(k_{nF}a_{nn})$ behave very differently in the low density
region. In \fig{fig7} and \fig{fig8}, the vertical straight line
located at $\r=0.15\r_0$ is determined by the condition
$\xi_{nn}/d_{nn}=1$, which indicates the BCS region at higher
density. However, one can not define the BEC region through the
definition $\m_n=0$, since $\m_n$ is always positive in the whole
density region, and correspondingly the scaled condensate
$\D_{nn}/\e_{nF}$ is still small and the effective scattering
length $1/(k_{nF}a_{nn})$ is still positive at extremely low
density. All of these characteristics indicate that no boson
degree of freedom emerges in neutron matter at low density.
%%%%%%%%%%%%%%%%%%%%%%%%%%%%%%%%%%%%%%%%%%%%%%%%%%%%%%%%%%%%%%%%%%%%%%%
%%%%%%%%%%%%%%%%%%%%%%%%%%%%%%%%%%%%%%%%%%%%%%%%%%%%%%%%%%%%%%%%%%%%%%%
%%%%%%%%%%%%%%%%%%%%%%%%%%%%%%%%%%%%%%%%%%%%%%%%%%%%%%%%%%%%%%%%%%%%%%%
%%%%%%%%%%%%%%%%%%%%%%%%%%%%%%%%%%%%%%%%%%%%%%%%%%%%%%%%%%%%%%%%%%%%%%%

One can understand the reason why there is no BEC in neutron
matter in terms of the density dependent coupling constant $g_I$
and the Fermi surface $k_{nF}$. For the DDCI potential parameters
chosen by fitting the pairing gap versus Fermi momentum, the
effective coupling in $I=1$ channel is weaker than the one in
$I=0$ channel at any density, $|g_1(\r)|<|g_0(\r)|$, and on the
other hand, the neutron density $\r_n=\r$ in neutron matter is two
times the neutron or proton density $\r_n = \r_p = \r/2$ in
symmetric matter. From this comparison, while BEC can form in
symmetric nuclear matter, its formation in a dense neutron matter
with weak coupling becomes difficult and even impossible. This
explains also why the BCS boundary shifts from $\r/\r_0 = 0.4$ in
symmetric matter to $0.15$ in neutron matter.

%%%%%%%%%%%%%%%%%%%%%%%%%%%%%%%%%%%%%%%%%%%%%%%%%%%%%%%%%%%%%%%%%%%%%%%
\subsection {Asymmetric Nuclear Matter: $0<\a<1$}\label{3.3}
%%%%%%%%%%%%%%%%%%%%%%%%%%%%%%%%%%%%%%%%%%%%%%%%%%%%%%%%%%%%%%%%%%%%%%%
While only $np$ condensate in symmetric nuclear matter with $\a=0$
and $nn$ condensate in neutron matter with $\a=1$ can survive,
there may exist $np,\ nn$ and $pp$ condensates in asymmetric
nuclear matter with $0<\a<1$. In this general case, the three
condensates $\D_{np},\ \D_{nn}$ and $\D_{pp}$ and the chemical
potentials $\m$ and $\delta\m$ as functions of nuclear density
$\r$ and asymmetry $\delta\r$ are calculated by solving the
coupled three gap equations and two number equations, and the
ground state of the system is determined by comparing the
corresponding free energies.
%%%%%%%%%%%%%%%%%%%%%%%%%%%%%%%%%%%%%%%%%%%%%%%%%%%%%%%%%%%%%%%%%%%%%%%
\begin{figure}[!htb]
\begin{center}
\includegraphics[width=8.5cm,height=7.5cm]{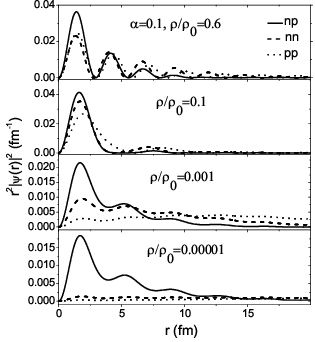}
\caption{ The probabilities $r^2|\j_{np}(r)|^2,\
r^2|\j_{nn}(r)|^2$ and $r^2|\j_{pp}(r)|^2$ as functions of the
relative distance $r$ between the pair partners at different
nuclear density in nuclear matter with asymmetry $\a=0.1$.}
\label{fig9}
\end{center}
\end{figure}
%%%%%%%%%%%%%%%%%%%%%%%%%%%%%%%%%%%%%%%%%%%%%%%%%%%%%%%%%%%%%%%%%%%%%%%

The three probabilities $r^2|\j_{np}(r)|^2,\ r^2|\j_{nn}(r)|^2$
and $r^2|\j_{pp}(r)|^2$ for $np,\ nn$ and $pp$ Cooper pairs are
shown in \fig{fig9} at different nuclear density and fixed
asymmetry $\a=0.1$. At high density $\r/\r_0=0.6$, all the
probabilities are spatially extended with strong Friedel
oscillations, indicating that all paired nucleons are weakly
correlated and the ground state is in the BCS phase with all three
condensates. Similar to the case in symmetric nuclear matter, with
decreasing nuclear density $\r$ the $np$ pairing probability
shrinks first but slightly expands again at extremely low density.
The very surprising feature is that the $nn$ and $pp$ pairing
probability are spatially expanded in a wide region at low density
and even approximately $r$-independent at extremely low density.
This is a strong hint that there are no BEC states of $nn$ and
$pp$ pairings in general asymmetric nuclear matter.
%%%%%%%%%%%%%%%%%%%%%%%%%%%%%%%%%%%%%%%%%%%%%%%%%%%%%%%%%%%%%%%%%%%%%%%
\begin{figure}[!htb]
\begin{center}
\includegraphics[width=7cm]{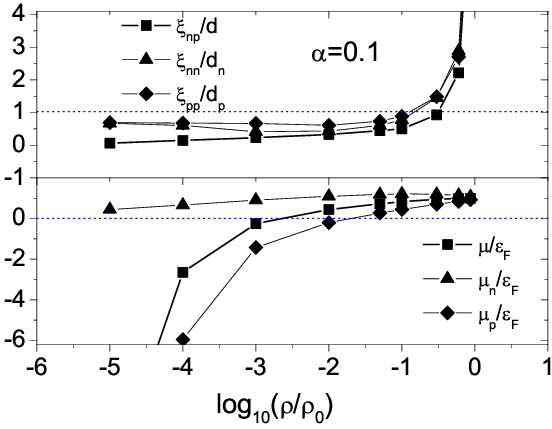}
\caption{ The scaled root-mean-square radius $\x_{ij}/d_{ij}$ and
scaled chemical potential $\m_i/\e_F$ as functions of nuclear
density in nuclear matter with asymmetry $\a=0.1$. }
\label{fig10}
\end{center}
\end{figure}
%%%%%%%%%%%%%%%%%%%%%%%%%%%%%%%%%%%%%%%%%%%%%%%%%%%%%%%%%%%%%%%%%%%%%%%

To check if the system could reach BEC state at low density, we
calculate the scaled root-mean-square radius $\x_{ij}/d_{ij}$ and
chemical potential $\m_i/\e_F$ and show them in \fig{fig10} as
functions of nuclear density at fixed asymmetry $\a=0.1$. The BCS
region for the $ij$ pairing is defined through the condition
$\x_{ij}/d_{ij}=1$. The BCS boundary is roughly at $\r/\r_0 =
0.35$ for $np$ pairs, $0.14$ for $pp$ pairs and $0.17$ for $nn$
pairs. Below the BCS boundary the system is regarded as a strongly
correlated superfluid. As mentioned above, for a system with only
one kind of pairings such as symmetric nuclear matter, the
strongly correlated system will go into the BEC state when the
chemical potential becomes negative. With this judgement, the $np$
Cooper pairs could reach BEC state at the critical density
$\r/\r_0 = 0.002$ where the chemical potential $\m$ changes sign.
For the $nn$ pairs, they could never form di-neutron bound state,
since the neutron chemical potential is always positive, although
it drops down as density decreases. The interesting phenomenon is
for the $pp$ pairing. At $\r/\r_0 = 0.03$, the proton chemical
potential changes sign, even earlier than the change for the $np$
pairing! Does the negative $\m_p$ here mean the formation of $pp$
BEC state? Is it inconsistent with the flat structure of the
probability shown in \fig{fig9}? To answer this questions, we
should note that for a general asymmetric system, the three
condensates are strongly coupled and for any anomalous density
$\j_{ij}(\bk)=\Delta_{ij}/(2E_{i\bk})$ there is no simple
Schr\"odinger-like equation \eq{schronp}, and the corresponding
chemical potential $\m_i$ at low density limit can not be
identified as a half of the binding energy of the di-nucleon bound
state. Therefore, there is no longer a definite relation between
the sign change of $\m_i$ and the BEC formation.

\begin{figure}[!htb]
\begin{center}
\includegraphics[width=7cm]{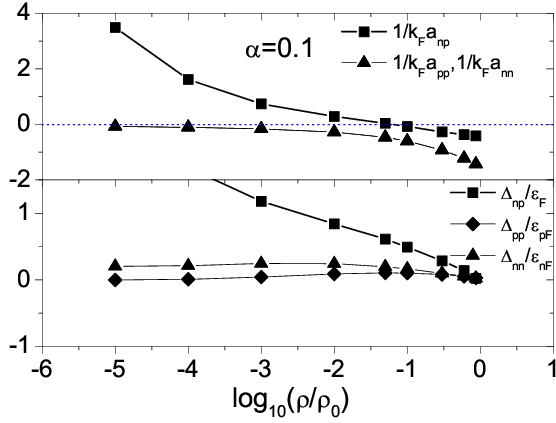}
\caption{ The scaled condensates $\D_{np}/\e_F$,
$\D_{nn}/\e_{nF}$, $\D_{pp}/\e_{pF}$ and scattering lengthes
$1/(k_Fa_{np})$, $1/(k_Fa_{nn})$, $1/(k_Fa_{pp})$ as functions of
nuclear density in nuclear matter with asymmetry $\a=0.1$.}
\label{fig11}
\end{center}
\end{figure}

\begin{figure}[!htb]
\begin{center}
\includegraphics[width=7cm]{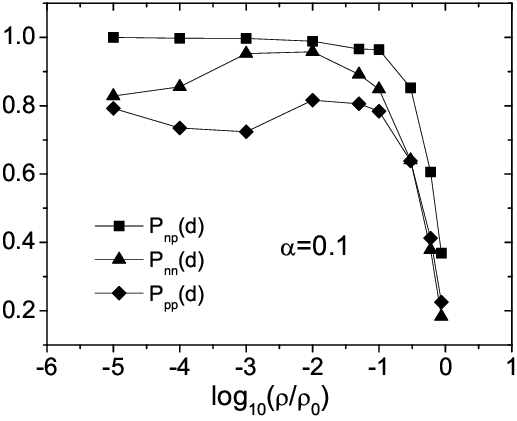}
\caption{ The probability $P_{ij}(d)$ as a function of nuclear
density in nuclear matter with asymmetry $\a=0.1$. }
\label{fig12}
\end{center}
\end{figure}

To verify whether the BEC state is reached, we calculate the
effective s-wave scattering length $1/(k_Fa_{ij})$, the scaled
condensate $\D_{ij}/\e_{iF}$, and the probability $P_{ij}(d)$ as
functions of nuclear density at fixed asymmetry $\a = 0.1$. The
results are presented in \fig{fig11} and \fig{fig12}. The
scattering length in $I=0$ channel changes its sign at about
$\r/\r_0 = 0.07$, but the other two lengthes in $I=1$ channel
remain negative in the whole density region. Correspondingly, the
scaled condensate $\D_{np}/\e_F$ becomes much larger than $1$ at
low density but $\D_{nn}/\e_{nF}$ and $\D_{pp}/\e_{pF}$ are still
very small even at extremely low density, and $P_{np}(d)$ is
almost equal to $1$ but $P_{nn}(d)$ and $P_{pp}(d)$ are clearly
less than $1$ at low density. Therefore, we can safely say that,
the $np$ BEC state is reached at low density, but there are no
$nn$ and $pp$ BEC states in general asymmetric nuclear matter.
%%%%%%%%%%%%%%%%%%%%%%%%%%%%%%%%%%%%%%%%%%%%%%%%%%%%%%%%%%%%%%%%%%%%%%%
%%%%%%%%%%%%%%%%%%%%%%%%%%%%%%%%%%%%%%%%%%%%%%%%%%%%%%%%%%%%%%%%%%%%%%%
%%%%%%%%%%%%%%%%%%%%%%%%%%%%%%%%%%%%%%%%%%%%%%%%%%%%%%%%%%%%%%%%%%%%%%%
%%%%%%%%%%%%%%%%%%%%%%%%%%%%%%%%%%%%%%%%%%%%%%%%%%%%%%%%%%%%%%%%%%%%%%%

The behavior of the nuclear superfluid in general case with
$0<\a<1$ is controlled by the two constraints $|g_1(\r)| <
|g_0(\r)|$ and $\r_p=(1-\a)/2\r < \r_n=(1+\a)/2\r < \r$. For $np$
pairings, the effective coupling constant $g_0$ is density
independent in our case, and the crossover from BCS to BEC is
characterized by the density dependent nucleon mass (\ref{m}). For
$nn$ or $pp$ pairings, the increase of the density leads to a high
Fermi surface and then makes the pairing easy, but on the other
hand the coupling constant drops down with increasing density. It
is the balance between these two effects that the probability
$P_{nn}(d)$ or $P_{pp}(d)$ is not a monotonous function of density
but there exists a valley structure at low density. Since the
coupling constant $g_1$ is the same for $nn$ and $pp$ pairings but
the Fermi surface for $nn$ pairing is higher than that for $pp$
pairing, $pp$ pairing is most impossible to form a BEC state, that
is the reason why the scaled condensate and probability for $pp$
pairing are less than the corresponding values for $nn$ pairing,
see \fig{fig11} and \fig{fig12}. When we consider Coulomb effect
which is neglected in our treatment and which will break the
relation $g_{nn}=g_{pp}$, the much stronger screening effect for
$pp$ interaction leads to $|g_{pp}|<|g_{nn}|$. As a result, $pp$
pairing will become further impossible to be in BEC state.

%%%%%%%%%%%%%%%%%%%%%%%%%%%%%%%%%%%%%%%%%%%%%%%%%%%%%%%%%%%%%%%%%%%%%%%
\subsection {Phase Diagram in $\r-\a$ Plane }\label{3.3}
\begin{figure}[!htb]
\begin{center}
\includegraphics[width=7cm]{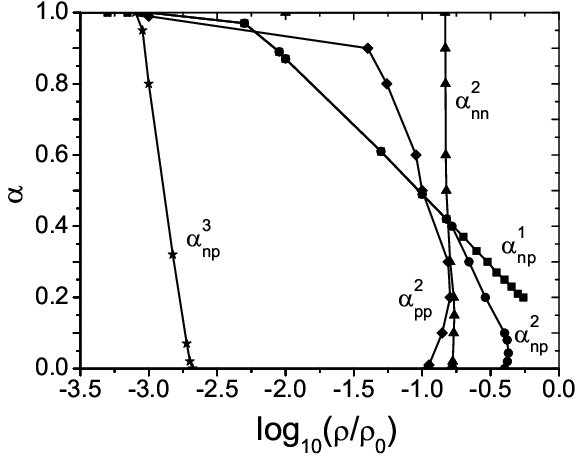}
\caption{ The phase diagram for asymmetric nuclear matter in
$\r-\a$ plane at zero temperature. $\a_{np}^1$ is the critical
asymmetry for $np$ pairing BCS superfluid, $\a_{np}^2$ and
$\a_{np}^3$ are the two boundaries of $np$ pairing BCS-BEC
crossover, and $a_{nn}^2$ ($\a_{pp}^2$) is the boundary for $nn$
($pp$) pairing to enter the strongly coupled region. }
\label{fig13}
\end{center}
\end{figure}

%%%%%%%%%%%%%%%%%%%%%%%%%%%%%%%%%%%%%%%%%%%%%%%%%%%%%%%%%%%%%%%%%%%%%%%
The phase diagram of asymmetric nuclear matter in the $\r-\a$
plane at zero temperature is presented in \fig{fig13}. For $\a=0$
only $\D_{np}$ could exist, and for $\a=1$ only $\D_{nn}$
survives. These two limits have been discussed in \ref{3.1} and
\ref{3.2}, and we will in this sub-section consider the phase
structure in the region of $0<\a<1$.

Due to the mismatch between neutron and proton Fermi surfaces
induced by the asymmetry $\a\neq 0$, there should be no $np$
pairing when the mismatch is large enough. Since the coupling
constant in $I=0$ channel is density independent and the pairing
is dynamically controlled only by the in-medium mass, the critical
asymmetry $\a_{np}^1(\rho)$ for the BCS superfluid increases with
decreasing density.  When the system enters the superfluid state,
the correlation becomes more and more strong with decreasing
density and the system will go into the crossover region at a
critical density defined by $\xi_{np}/d=1$. When the density
further decreases, the $np$ pairing starts to be in the BEC state
at another critical density defined by $\m=0$. Corresponding to
the above two critical densities, the two boundaries of the
BCS-BEC crossover in \fig{fig13} are indicated respectively by
$\alpha_{np}^2(\rho)$ and $\alpha_{np}^3(\rho)$. At higher
asymmetry, $\a_{np}^1$ and $\a_{np}^2$ coincide and the nuclear
matter goes directly from the normal state into the strongly
correlated superfluid.

Different from the $I=0$ channel, the pairing in $I=1$ channel has
no Fermi surface mismatch and the superfluid can survive at any
asymmetry when the density is not high enough. In \fig{fig13} the
$I=1$ superfluid exists in the whole $\r-\a$ plane. As we
mentioned above, at $\a=0.1$ the critical density for $nn$ pairing
to go into the strongly coupled region is larger than the one for
$pp$ pairing. This is true for any asymmetry. In \fig{fig13} the
boundary $\alpha_{nn}^2(\rho)$ defined by $\xi_{nn}/d_n = 1$ for
$nn$ pairing is always on the right-hand side of the one
$\alpha_{pp}^2(\rho)$ defined by $\xi_{pp}/d_p=1$ for $pp$
pairing. It is interesting that the critical density for $nn$
pairing is almost a constant for any asymmetry. There is no room
for the BEC state of $nn$ or $pp$ pairing at any asymmetry $0\leq
\a \leq 1$.

Note that there exist three jumps in the phase diagram. The jump
from normal state at $\r=0$ to the $np$ pairing BEC state at
$\r\neq 0$, the jump from $np$ pairing at $\a=0$ to the $np,\ nn$
and $pp$ pairings at $\a\neq 0$, and the jump from $nn$ pairing at
$\a=1$ to $np,\ nn$ and $pp$ pairings at $\a < 1$.
%%%%%%%%%%%%%%%%%%%%%%%%%%%%%%%%%%%%%%%%%%%%%%%%%%%%%%%%%%%%%%%%%%%%%%%
%%%%%%%%%%%%%%%%%%%%%%%%%%%%%%%%%%%%%%%%%%%%%%%%%%%%%%%%%%%%%%%%%%%%%%%

%%%%%%%%%%%%%%%%%%%%%%%%%%%%%%%%%%%%%%%%%%%%%%%%%%%%%%%%%%%%%%%%%%%%%%%
\section {Thermodynamics at High density}\label{4}
%%%%%%%%%%%%%%%%%%%%%%%%%%%%%%%%%%%%%%%%%%%%%%%%%%%%%%%%%%%%%%%%%%%%%%%
The Cooper pairs could be broken by both large mismatch between
the Fermi surfaces of the partner nucleons and large thermal
fluctuation. The former is characterized by the asymmetry $\a$,
and the latter is described by temperature $T$. We in this section
examine the thermal and asymmetric effect on the nuclear matter in
weak coupling region, $i.e.$, in the BCS region where the mean
field analysis is still reliable. From the above calculation at
zero temperature, this region is around and below the normal
nuclear density $\r_0$. Our purpose is to determine the phase
diagram in $\a-T$ plane at a fixed density.

We must emphasize that the asymmetry may lead to the so-called
phase separating instability, which means that the ground state
may favor the spatial separation of different phases. To study
such an instability we define the nuclear density susceptibility
matrix $\c$ with elements $\c_{ij}=\pt\m_i/\pt\r_j$ with
$i,j=n,p$. For an uniform matter, $\c$ is always positively
defined, and $\c<0$ can be used as an indication of the appearance
of phase separation (One should note that $\c<0$ is only a
necessary but not sufficient condition for the occurrence of phase
separation). In the following discussion the term phase separation
is refer to the state with $\c<0$. In a practical manner, $\c$ can
be computed through~\cite{Huang:2007}
\begin{eqnarray}
\c_{ij}&=&-{\pt^2\cf\over\pt \r_i\pt \r_j}\bigg|_\r\non
&=&-{\pt^2\O\over\pt\m_i\pt\m_j}\bigg|_\m +Y_i R^{-1}Y^\dag_j
\end{eqnarray}
with \begin{widetext}
\begin{eqnarray}
Y_i&=&\lb{\pt^2\O\over\pt\m_i\pt\D_{nn}}\bigg|_\m,
{\pt^2\O\over\pt\m_i\pt\D_{pp}}\bigg|_\m, {\pt^2\O\over\pt\m_i\pt
\D_{np}}\bigg|_\m, {\pt^2\O\over\pt\m_i\pt q}\bigg|_\m\rb,\non
R&=&\lb\begin{array}{cccc} {\pt^2\O\over\pt \D_{nn}^2} &
{\pt^2\O\over\pt \D_{nn}\pt\D_{pp}} &
{\pt^2\O\over\pt\D_{nn}\pt\D_{np}} & {\pt^2\O\over\pt\D_{nn}\pt q}\\
{\pt^2\O\over\pt \D_{pp}\pt\D_{nn}} & {\pt^2\O\over\pt\D_{pp}^2} &
{\pt^2\O\over\pt\D_{pp}\pt\D_{np}} & {\pt^2\O\over\pt\D_{pp}\pt q}\\
{\pt^2\O\over\pt \D_{np}\pt\D_{nn}} &
{\pt^2\O\over\pt\D_{np}\pt\D_{pp}} & {\pt^2\O\over\pt\D_{np}^2} &
{\pt^2\O\over\pt\D_{np}\pt q}\\
{\pt^2\O\over\pt q\pt \D_{nn}} & {\pt^2\O\over\pt q\pt\D_{pp}} &
{\pt^2\O\over\pt q\pt\D_{np}} & {\pt^2\O\over\pt
q^2}\end{array}\rb.
\end{eqnarray}
\end{widetext}

%%%%%%%%%%%%%%%%%%%%%%%%%%%%%%%%%%%%%%%%%%%%%%%%%%%%%%%%%%%%%%%%%%%%%%%
\begin{figure}[!htb]
\begin{center}
\includegraphics[width=8cm]{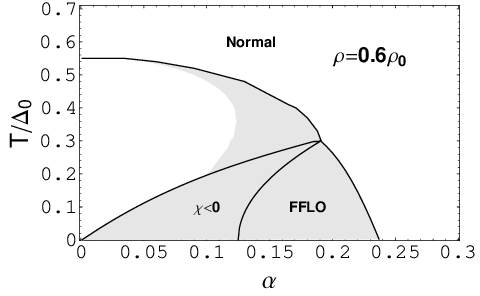}
\includegraphics[width=8cm]{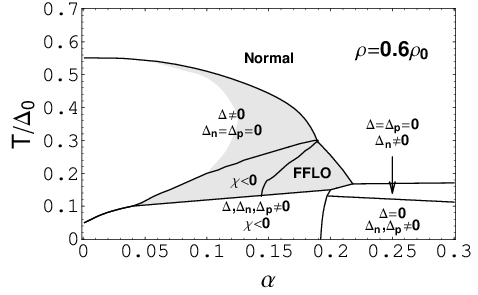}
\caption{ The phase diagram of a mismatched nuclear superfluid at
fixed nuclear density$\r/\r_0=0.6$ in $\a-T$ plane. The upper
panel is for the familiar superfluid with only $I=0$ pairing, and
in the lower panel the $I=1$ pairing is included as well. The
shadowed regions indicate the gapless superfluid. $\D_0 = 7.7$ MeV
is the condensate in $I=0$ channel at $\a=0$ and $T=0$. }
\label{fig14}
\end{center}
\end{figure}
%%%%%%%%%%%%%%%%%%%%%%%%%%%%%%%%%%%%%%%%%%%%%%%%%%%%%%%%%%%%%%%%%%%%%%%
By calculating the coupled gap and number equations at finite
temperature, we obtain all the possible homogeneous phases and
inhomogeneous FFLO phase. From the comparison of their free
energies, we then extract the lowest one at fixed $\r,\ \alpha$
and $T$. Finally we investigate the stability of the system
against the number fluctuations by computing the number
susceptibility matrix $\chi$, the state with non positive-definite
$\chi$ may be a phase separation. In \fig{fig14} we show the phase
diagram in $\a-T$ plane at fixed nuclear density $\r/\r_0=0.6$.

The phase diagram without $I=1$ pairing is presented in the upper
panel. The system is in normal state when the asymmetry or
temperature is high enough, and the superfluid is in homogeneous
state at high temperature and FFLO state at low temperature.
However, the number susceptibility in the region of low
temperature and low number asymmetry is not positive-definite, the
FFLO state in this region is therefore unstable against the number
fluctuations, and the ground state is probably an inhomogeneous
mixture of the BCS superfluid and normal nuclear fluid. The
shadowed region is the gapless superfluid with $\delta\mu >
\Delta_{np}$ where the energy gap to excite quasi-nucleons is zero
and the system may be sensitive to the thermal and quantum
fluctuations.

The phase diagram with both $I=0$ and $I=1$ pairings is shown in
the lower panel of \fig{fig14}. Besides the familiar phase with
only $I=0$ pairing ($\Delta_{np}\ne 0, \Delta_{nn}=\Delta_{pp}=0$)
and the expected phases with only $I=1$ pairing ($\Delta_{nn},
\Delta_{pp}\ne 0, \Delta=0$ and $\Delta_{nn}\ne 0,
\Delta_{np}=\Delta_{pp}=0$), there appears a new phase where the
two kinds of pairings coexist ($\Delta_{np}, \Delta_{nn},
\Delta_{pp}\ne 0$). In this new phase the FFLO momentum is zero
and the number susceptibility is negative, $\chi < 0$. Therefore,
the homogeneous superfluid in this region is unstable against
number fluctuations, and the ground state is probably an
inhomogeneous mixture of these three superfluid components. In the
familiar phase with only $I=0$ pairing, there remains a stable
FFLO region and an unstable FFLO triangle where the number
susceptibility is negative and the system may be in the phase
separation of the BCS superfluid and normal nuclear fluid. The
gapless state appears only in the $I=0$ pairing superfluid, and in
the region with $I=1$ pairing all the nucleons are fully gapped.
Since the Fermi surface for $nn$ pairing is higher than the one
for $pp$ pairing, the critical temperature to melt the $nn$
condensate is higher than that for melting the $pp$ condensate,
and the difference between the two increases with increasing
asymmetry.

%%%%%%%%%%%%%%%%%%%%%%%%%%%%%%%%%%%%%%%%%%%%%%%%%%%%%%%%%%%%%%%%%%%%%%
\section {Summary}\label{5}
%%%%%%%%%%%%%%%%%%%%%%%%%%%%%%%%%%%%%%%%%%%%%%%%%%%%%%%%%%%%%%%%%%%%%%
We have investigated the phase structure of isospin asymmetric
nuclear superfluid with pairings in both $I=0$ and $I=1$ channels
in the frame of the density dependent contact potential. We
calculated at zero temperature and in mean field approximation the
pair wave functions, pair condensates, nucleon chemical potentials
and effective couplings which are normally considered as
characteristic quantities describing BCS-BEC crossover, and found
that in general asymmetric nuclear matter only the $np$ pair could
form true bound state at extremely low density, and $nn$ and $pp$
pairs, on the other hand, could never form bound state at any
density and asymmetry. We also studied the phase diagram for
weakly coupled nuclear superfluid at finite temperature. Since the
attractive interaction for $I=1$ pairing is weaker than the one
for $I=0$ pairing, the inclusion of $nn$ and $pp$ pairings changes
significantly the conventional phase diagram with only $np$
pairing only at low temperature. For systems with fixed nuclear
density, the two kinds of pairings can coexist at low temperature
and low number asymmetry. By calculating the number
susceptibility, this new phase is not in the FFLO state but
probably an inhomogeneous mixture of the $np,\ nn$ and $pp$
superfluid components. In any region with $I=1$ pairing, the
interesting gapless superfluid is washed out and all
quasi-nucleons are fully gapped.

{\bf Acknowledgments:} X.H. thanks A.Sedrakian for helpful
discussions. The work is supported by the NSFC Grant 10735040 and
the National Research Program Grants 2006CB921404 and
2007CB815000.

\end{document}